\documentclass[11pt]{cernrep}
\usepackage{graphicx}
\usepackage{here}
\usepackage{amssymb,amsfonts}
\begin{document}
\title{Enhancing the physical significance of Frequentist confidence intervals\footnote{
Talk preseted at the Workshop on ``Confidence Limits'',
CERN, 17-18 January 2000.
DFTT 07/00, arXiv:hep-ex/0002042.}}
\author{Carlo Giunti}
\institute{INFN, Sezione di Torino, and Dipartimento di Fisica Teorica,
Universit\`a di Torino,
Via P. Giuria 1, I--10125 Torino, Italy}
\maketitle
\begin{abstract}
It is shown that all the Frequentist methods are equivalent from a
statistical point of view,
but the physical significance of the confidence intervals
depends on the method.
The Bayesian Ordering method is
presented and confronted with the Unified Approach in the
case of a Poisson process with background.
Some criticisms to
both methods are answered.
It is also argued that a general Frequentist method is not needed.
\end{abstract}

\section{Introduction}
\label{Introduction}

In this report I will be concerned mainly with
the Frequentist (classical) theory of
statistical inference,
but I think that it is interesting and useful that
I express my opinion on
the war between Frequentists and Bayesians.
To the question
\begin{center}
``Are you Frequentist or Bayesian''?
\end{center}
I answer
\begin{center}
``I like statistics.''
\end{center}
I think that if one likes statistics,
one can appreciate the beauty
of both Frequentist and Bayesian theories
and the subtleties involved in their formulation and application.
I think that both approaches are valid
from a statistical as well as physical point of view.
Their difference arises from different definitions of probability
and their results answer different statistical questions.
One can like more one of the two theories,
but I think that it is unreasonable
to claim that only one of them is correct,
as some partisans of that theory claim.
These partisans often produce
examples in which the other approach is shown to yield misleading or
paradoxical results.
I think that each theory should be appreciated and used
in its limited range of validity,
in order to answer the appropriate questions.
Finding some example in which one approach fails
does not disprove its correctness in many other cases
that lie in its range of validity.

My impression is that the Bayesian theory
(see, for example, \cite{DAgostini-YR3-99})
has a wider range of validity
because it can be applied to cases in which
the experiment can be done only once or a few times
(for example,
our thoughts in everyday decisions and judgments seem to
follow an approximate Bayesian method).
In these cases the Bayesian definition of probability
as \emph{degree of believe} seems to me the only one that makes sense
and is able to provide meaningful results.

Let me remind that
since Galileo an accepted basis
of scientific research is the
\emph{repeatability of experiments}.
This assumption justifies the Frequentist definition of probability
as ratio of the number of positive cases and
total number of trials in a large ensemble.
The concept of \emph{coverage} follows immediately:
a $100\alpha\%$ \emph{confidence interval} for a physical quantity $\mu$
is an interval that
contains (covers) the unknown true value of that quantity
with a Frequentist probability $\alpha$.
In other words,
a $100\alpha\%$ confidence interval for $\mu$
belongs to a set of confidence intervals
that can be obtained with a large ensemble of experiments,
$100\alpha\%$ of which contain the true value of $\mu$.

\section{The statistical and physical significance of confidence intervals}
\label{significance}

I think that
in order to fully appreciate the meaning and usefulness of
Frequentist confidence intervals
obtained with Neyman's method
\cite{Neyman-37,Eadie-71},
it is important to understand that
the experiments in the ensemble do not need to be identical,
as often stated,
or even similar,
but can be real, different experiments
\cite{Neyman-37,Cousins-95}.
One can understand this property in a simple way
\cite{Giunti-back-99}
by considering,
for example,
two different experiments that measure the same physical quantity $\mu$.
The $100\alpha\%$
classical confidence interval obtained from the results of each experiment
belongs by construction to a set of confidence intervals which can be obtained
with an ensemble of identical experiments
and contain the true value of $\mu$ with probability $\alpha$.
It is clear that the sum of these two sets of confidence intervals,
containing the two confidence intervals
obtained in the two different experiments,
is still a set of confidence intervals that contain the true
value of $\mu$ with probability $\alpha$.

Moreover,
for the same reasons
it is clear that
\emph{the results of different experiments
can also be analyzed with different Frequentist methods}
\cite{Giunti-Laveder-00},
\textit{i.e.} methods with correct coverage
but different prescriptions for the construction
of the confidence belt.
This for me is amazing and beautiful:
\textit{whatever method you choose you get a result
that can be compared meaningfully with the results
obtained by different experiments using different methods}!
It is important to realize, however,
that the choice of the Frequentist method must be done
independently of the knowledge of the data
(before looking at the data),
otherwise the property of coverage is lost,
as in the ``flip-flop'' example in Ref.~\cite{Feldman-Cousins-98}.

This property allow us to solve an apparent paradox
that follows from the recent proliferation
of proposed Frequentist methods
\cite{Feldman-Cousins-98,Giunti-bo-99,Ciampolillo-98,Roe-Woodroofe-99,%
Mandelkern-Schultz-99,Punzi-99}.
This proliferation seems to introduce a large degree of
subjectivity in the Frequentist approach,
supposed to be objective,
due to the need to choose one specific prescription
for the construction of the confidence belt,
among several available with similar properties.
From the property above,
we see that whatever Frequentist method one chooses,
if implemented correctly,
the resulting confidence interval
can be compared statistically
with the confidence intervals of other experiments
obtained with other Frequentist methods.
Therefore,
\emph{the subjective choice of a specific Frequentist method
does not have any effect from a statistical point of view}!

Then you should ask me:
\begin{center}
Why are you proposing a specific Frequentist method?
\end{center}
The answer lies in \emph{physics}, not statistics.
It is well known that the statistical analysis of the same data
with different Frequentist methods produce
different confidence intervals.
This difference is sometimes
crucial for the physical interpretation of the result of the experiment
(see, for example, \cite{Giunti-bo-99,Roe-Woodroofe-99}).
Hence,
the physical significance
of the confidence intervals obtained with different
Frequentist methods is sometimes
crucially different.
In other words,
\emph{the Frequentist method suffers
from a degree of subjectivity
from a physical, not statistical, point of view}.

\section{The beauty of the Unified Approach and its pitfalls}
\label{beauty}

The possibility to apply successfully Frequentist statistics
to problematic cases in frontier research has received
a fundamental contribution with the proposal of the Unified Approach
by Feldman and Cousins
\cite{Feldman-Cousins-98}.
The Unified Approach
consists in a clever prescription for the construction
of ``a classical confidence belt
which unifies the treatment of upper confidence limits for null results
and two-sided confidence intervals for non-null results".

In the following I will consider the case of
a Poisson process with signal $\mu$ and known background $b$.
The probability to observe $n$ events is
\begin{equation}
P(n|\mu,b)
=
\frac{ (\mu+b)^n e^{-(\mu+b)} }{ n! }
\,.
\label{Poisson}
\end{equation}

The Unified Approach is based on the construction
of the acceptance intervals
$[n_1(\mu),n_2(\mu)]$
ordering the $n$'s through their rank
given by the relative magnitude
of the likelihood ratio
\begin{equation}
R(n,\mu,b)
=
\frac{ P(n|\mu,b) }{ P(n|\mu_{\mathrm{best}},b) }
=
\left( \frac{\mu+b}{\mu_{\mathrm{best}}+b} \right)^n
e^{\mu_{\mathrm{best}}-\mu}
\,,
\label{LR}
\end{equation}
where $\mu_{\mathrm{best}}$
is the maximum likelihood estimate of $\mu$,
\begin{equation}
\mu_{\mathrm{best}}(n,b)
=
\mathrm{Max}[0,n-b]
\,.
\label{mu-best}
\end{equation}
As a result of this construction
the confidence intervals are two-sided
(\textit{i.e.} $[\mu_{\mathrm{low}},\mu_{\mathrm{up}}]$
with $ \mu_{\mathrm{low}} > 0 $)
for $n \gtrsim b$,
whereas for $n \lesssim b$ they are upper limits
(\textit{i.e.} $ \mu_{\mathrm{low}} = 0 $).

\begin{figure}[t]
\begin{minipage}[t]{0.48\textwidth}
\mbox{\includegraphics[bb=90 525 307 737,width=\textwidth]{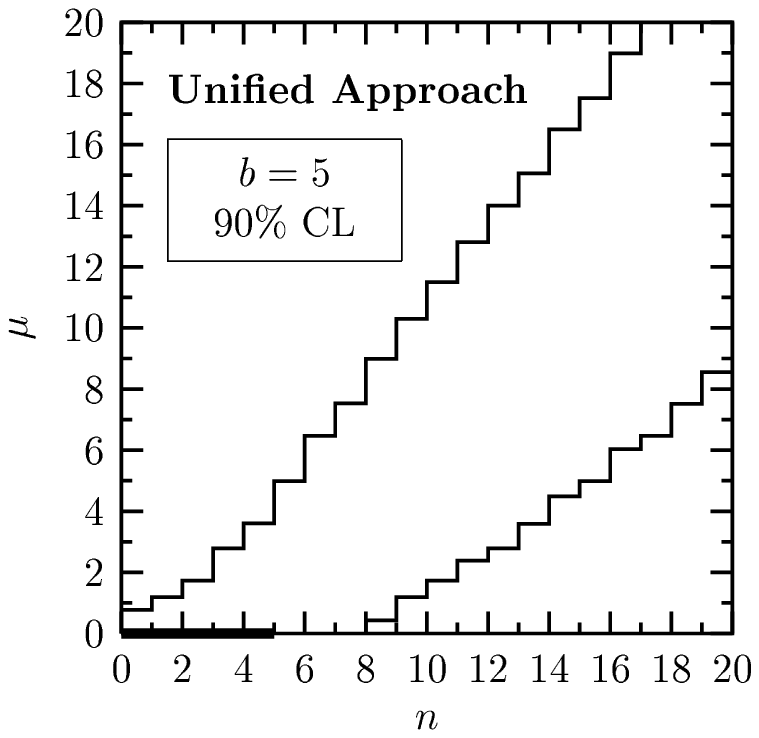}}
\caption{ \label{ualim}
Confidence belt in the Unified Approach
for background $b=5$ and confidence level $\alpha=0.90$.}
\end{minipage}
\hfill
\begin{minipage}[t]{0.48\textwidth}
\mbox{\includegraphics[bb=85 525 303 737,width=\textwidth]{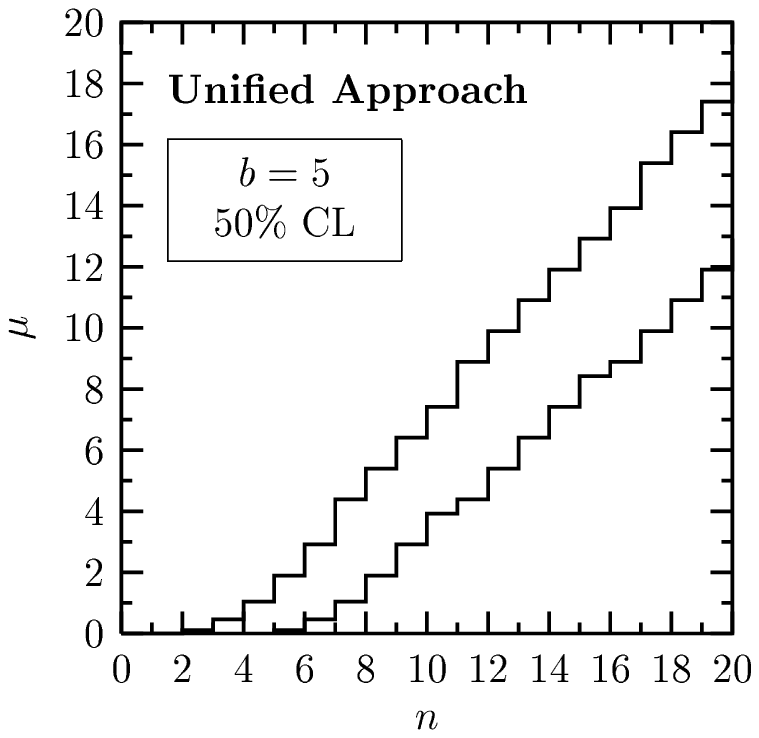}}
\caption{ \label{uabelt}
Confidence belt in the Unified Approach
for background $b=5$ and confidence level $\alpha=0.50$.}
\end{minipage}
\end{figure}

The fact that the confidence intervals are two-sided
for $n \gtrsim b$
can be understood by considering
$n > b$,
that gives
$\mu_{\mathrm{best}} = n - b$.
In this case
the likelihood ratio (\ref{LR}) is given by
\begin{equation}
R(n>b,\mu,b)
=
\left(\frac{\mu+b}{n}\right)^n
e^{n-(\mu+b)}
=
\exp\left\{
n \left[ 1 + \ln(\mu+b) - \ln{n} \right]
- (\mu+b)
\right\}
\stackrel{n\to\infty}{\longrightarrow}
0
\,.
\label{LR-2}
\end{equation}
This implies that the rank of high values of $n$ is very low
and they are excluded form the confidence belt.
Therefore, the acceptance intervals
$[n_1(\mu),n_2(\mu)]$
are always bounded,
\textit{i.e.} $n_2(\mu)$ is finite,
and the confidence intervals are two-sided
for $n \gtrsim b$,
as illustrated in Fig.~\ref{ualim},
where the solid lines show the borders of the confidence belt
for a background $b=5$ and a confidence level $\alpha=0.90$.

The fact that the confidence intervals are upper limits for
$n \lesssim b$
can be understood by considering
$n \leq b$,
for which we have
$\mu_{\mathrm{best}}=0$
and
the likelihood ratio that determines
the ordering of the $n$'s in the acceptance intervals is given by
\begin{equation}
R(n{\leq}b,\mu,b)
=
\left( 1 + \frac{\mu}{b} \right)^n
\,
e^{-\mu}
\,.
\label{LR-1}
\end{equation}
Considering now the acceptance interval
for $\mu = 0$,
we have
$R(n \leq b,\mu=0,b) = 1$.
Therefore,
all $n \leq b$
for $\mu = 0$
have highest rank and
are guaranteed to lie in the confidence belt.
This is illustrated in Fig.~\ref{ualim},
where
the thick solid segment
shows the $n \leq b$ part of the acceptance interval
for $\mu=0$,
that must lie in the confidence belt.
Since $\mu$ is a continuous parameter,
also for small values of $\mu$ the $n \leq b$
have rank close to the highest one
and lie in the confidence belt.
Indeed, for $\mu>0$,
the likelihood ratio (\ref{LR}) increases for $n$
going form zero to the largest integer smaller or equal to $b$
and decreases for larger values of $n$.
Hence,
the largest integer $n_{\mathrm{hr}}$
such that
$n_{\mathrm{hr}} \leq b$ has highest rank.
If $\mu$ is sufficiently small
all $n \leq b$ have rank close to maximum
and are included in the confidence belt if
the confidence level is large enough,
$\alpha \gtrsim 0.60$.
For example,
$R(n=0,\mu,b) > R(n_{\mathrm{hr}}+1,\mu,b)$
for
$\mu < (1+b) e^{-1/(1+b)} - b$.
Therefore,
the left edge of the confidence belt
must change its slope for $n \lesssim b$
and intercept the $\mu$-axis at a positive value of $\mu$,
as illustrated in Fig.~\ref{ualim}.
The value of $\mu$ at which the left edge of the confidence belt
intercepts the $\mu$-axis,
that corresponds to $\mu_{\mathrm{up}}(n=0)$,
depends on the value of the background $b$
and on the value of the confidence level $\alpha$.

However\footnote{
Let me emphasize that I discuss this case only for the sake
of curiosity.
It is pretty obvious that a low value of $\alpha$
is devoid of any practical interest.
},
for small values of $\alpha$
the Unified Approach
gives zero-width confidence intervals for $n \ll b$,
as illustrated in Fig.~\ref{uabelt},
where I have chosen $b=5$ and $\alpha=0.50$.
One can see that the segment
$n \leq b$ is enclosed in the confidence belt
for $\mu=0$,
but for any value of $\mu>0$
the sum of the probabilities of the $n$'s close to $\mu+b$
is enough to reach the confidence level
and low values of $n$ are not included in the confidence belt.
Hence,
in this case the Unified Approach gives zero-width
confidence intervals for $n<2$.

The unification of the
treatments of upper confidence limits for null results
and two-sided confidence intervals for non-null results
obtained with
the Unified Approach is wonderful,
but it has been noticed that
the upper limits
obtained with the Unified Approach
for $n < b$
are too stringent (meaningless)
from a physical point of view
\cite{Giunti-bo-99,Giunti-98}.
In other words,
although these limits are statistically correct
from a Frequentist point of view,
they cannot be taken as reliable upper bounds
to be used in physical applications.

\begin{figure}[t]
\mbox{\includegraphics[bb=54 491 517 662,width=\textwidth]{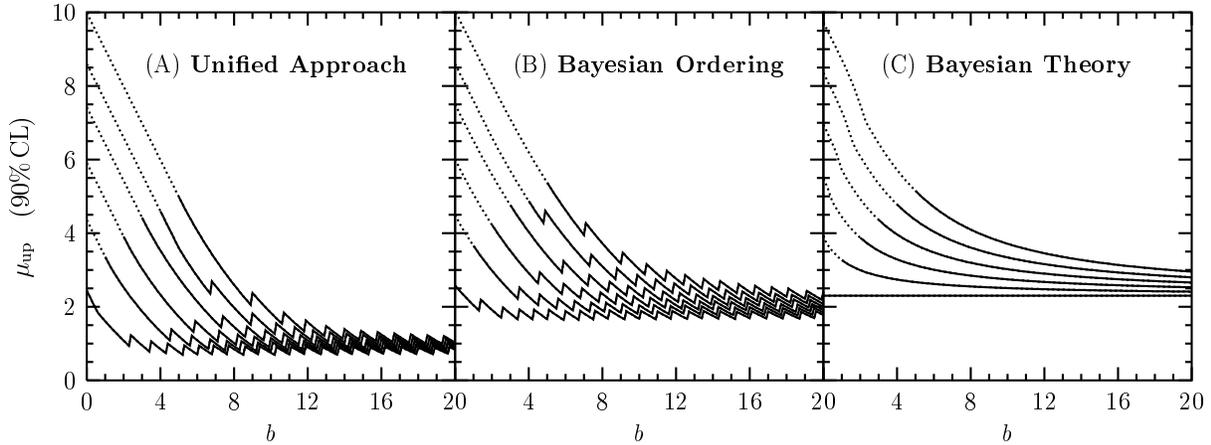}}
\caption{ \label{higback}
90\% CL upper limit
$\mu_{\mathrm{up}}$
as a function of the background $b$
for $n = 0$ (lower lines),
$\ldots$,
$n = 5$ (upper lines).
The solid part of each line shows where $b \geq n$.
}
\end{figure}

This problem is illustrated in Fig.~\ref{higback}A,
where I plotted the 90\% CL upper limit
$\mu_{\mathrm{up}}$
as a function of $b$
for $n = 0, \ldots, 5$.
The solid part of each line shows where $b \geq n$.
One can see that for a given $n$,
$\mu_{\mathrm{up}}$
decreases rather steeply when $b$ is increased,
until a minimum value close to one is reached.
The curves have jumps
because $n$ is an integer and generally the desired confidence level
cannot be obtained exactly,
but with some unavoidable overcoverage.

Let me emphasize that the problem of obtaining too stringent
upper limits for $n < b$ is very serious for
a scientist that wants to obtain reliable information from experiment
and use this information
for other purposes
(as input for a theory or another experiment).
In the past,
researchers
bearing the same physical point of view
refrained to report
empty confidence intervals
or very stringent upper limits
when $n < b$ was measured.
These confidence intervals are correct from a statistical point of view,
but useless from a physical point of view.
Furthermore,
the same reasoning lead to prefer
the Unified Approach to central confidence intervals
or upper limits,
because the non-empty
confidence interval
obtained when $n < b$ is measured
is certainly more significant,
from a physical point of view,
than an empty one,
although they are statistically equivalent,
as shown in Section~\ref{significance}.

\section{A brutal modification of the Unified Approach}
\label{Brutal}

In the Unified Approach
$\mu_{\mathrm{best}}$
is positive and equal to zero for $n \leq b$.
If
$\mu_{\mathrm{best}}$
is forced to be always bigger than zero,
the $n$'s smaller than $b$ have rank higher than in the Unified Approach.
As a consequence,
the decrease of the upper limit
$\mu_{\mathrm{up}}$
as $b$ increases is weakened.
This is illustrated by a
\emph{``Brutally Modified Unified Approach''}
(BMUA)
in which we take
\begin{equation}
\mu_{\mathrm{best}}
=
\mathrm{Max}[\mu_{\mathrm{best}}^{\mathrm{min}},n-b]
\,,
\label{mu-best-BMUA}
\end{equation}
where
$\mu_{\mathrm{best}}^{\mathrm{min}}$
is a positive real number.

\begin{figure}[t]
\begin{minipage}[t]{0.48\textwidth}
\mbox{\includegraphics[bb=90 525 307 737,width=\textwidth]{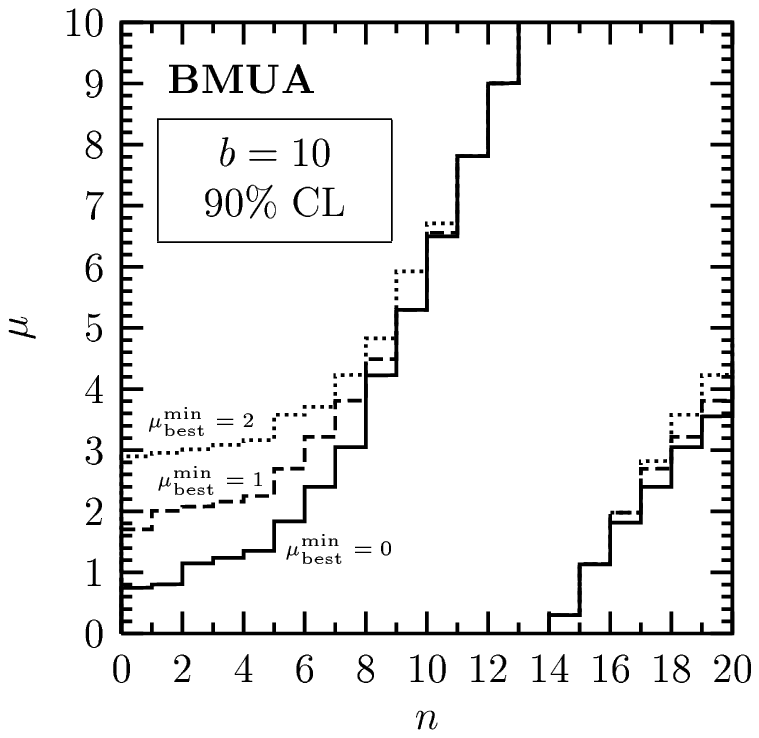}}
\caption{ \label{aubelt}
90\% confidence belts for $b=10$
in the Unified Approach
($\mu_{\mathrm{best}}^{\mathrm{min}} = 0$,
solid lines)
and in the Brutally Modified Unified Approach
(BMUA)
for
$\mu_{\mathrm{best}}^{\mathrm{min}} = 1$
(dashed lines)
and
$\mu_{\mathrm{best}}^{\mathrm{min}} = 2$
(dotted lines).
}
\end{minipage}
\hfill
\begin{minipage}[t]{0.48\textwidth}
\mbox{\includegraphics[bb=85 525 303 737,width=\textwidth]{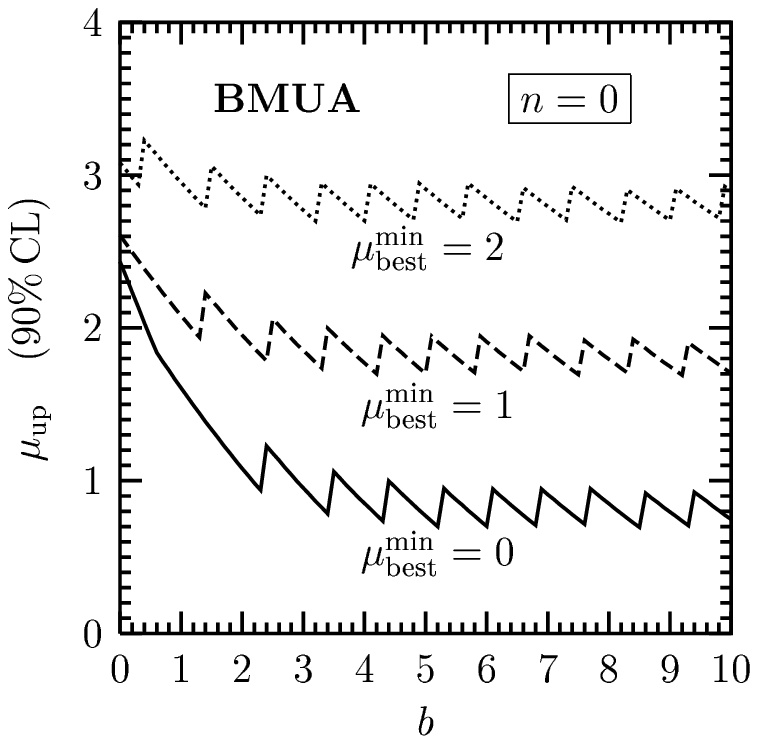}}
\caption{ \label{auhigback}
90\% CL upper limit
$\mu_{\mathrm{up}}$
as a function of the background $b$
for $n = 0$
in the Unified Approach
($\mu_{\mathrm{best}}^{\mathrm{min}} = 0$,
solid line)
and in the BMUA
for
$\mu_{\mathrm{best}}^{\mathrm{min}} = 1$
(dashed line)
and
$\mu_{\mathrm{best}}^{\mathrm{min}} = 2$
(dotted line).
}
\end{minipage}
\end{figure}

In Fig.~\ref{aubelt}
I plotted the confidence belts for
$\mu_{\mathrm{best}}^{\mathrm{min}} = 0$
(solid lines),
that corresponds to the Unified Approach,
$\mu_{\mathrm{best}}^{\mathrm{min}} = 1$
(dashed lines)
and
$\mu_{\mathrm{best}}^{\mathrm{min}} = 2$
(dotted lines),
for $b=10$.
One can see that in the BMUA
the upper limits of the confidence intervals
are considerably higher than in the Unified Approach.
The behavior of
$\mu_{\mathrm{up}}$
as a function of $b$ for $n=0$
is shown in Fig.~\ref{auhigback},
from which it is clear that the decrease of
$\mu_{\mathrm{up}}$
when $b$ increases is much weaker in the BMUA
(dashed and dotted lines)
than in the Unified Approach
(solid line)
and it is almost absent for
$\mu_{\mathrm{best}}^{\mathrm{min}} \gtrsim 2$.

Let me emphasize that

\begin{enumerate}

\item
The BMUA is a statistically correct
Frequentist method and coverage is satisfied.

\item
In the BMUA one obtains upper limits for $n \lesssim b$
and central confidence intervals for $n \gtrsim b$,
as in the Unified Approach\footnote{
For
$n \leq b + \mu_{\mathrm{best}}^{\mathrm{min}}$
we have
$\mu_{\mathrm{best}} = \mu_{\mathrm{best}}^{\mathrm{min}}$
and the likelihood ratio (\ref{LR}) becomes
\begin{equation}
R(n \leq b + \mu_{\mathrm{best}}^{\mathrm{min}},\mu,b)
=
\left( \frac{\mu+b}{\mu_{\mathrm{best}}^{\mathrm{min}}+b} \right)^n
e^{\mu_{\mathrm{best}}^{\mathrm{min}}-\mu}
\,.
\label{LR-11}
\end{equation}
For
$\mu<\mu_{\mathrm{best}}^{\mathrm{min}}$,
we have
$(\mu+b)/(\mu_{\mathrm{best}}^{\mathrm{min}}+b)<1$
and
$R(n \leq b + \mu_{\mathrm{best}}^{\mathrm{min}},\mu,b)$
decreases with increasing $n$.
Let us consider now
$n > b + \mu_{\mathrm{best}}^{\mathrm{min}}$,
for which
$\mu_{\mathrm{best}} = n - b$
and the likelihood ratio (\ref{LR}) is given by
the expression in Eq.~(\ref{LR-2}).
This expression has a maximum for $n$ equal to
one of the two integers
closest to $\mu+b$.
For
$\mu<\mu_{\mathrm{best}}^{\mathrm{min}}$,
this integer is the first one in the considered range
($n > b + \mu_{\mathrm{best}}^{\mathrm{min}}$).
Therefore,
for sufficiently low values of $\mu$,
$\mu<\mu_{\mathrm{best}}^{\mathrm{min}}$,
the likelihood ratio (\ref{LR})
decreases monotonically
as $n$ increases.
In this case,
low values of $n$ have highest ranks
and are guaranteed to lie in the confidence belt
and
the left edge of the confidence belt
must change its slope for
$n \lesssim \mu_{\mathrm{best}}^{\mathrm{min}} + b$
and intercept the $\mu$-axis at a positive value of $\mu$,
as illustrated in Fig.~\ref{aubelt}.
}.

\item
The BMUA method is not general
(although it can be extended in an obvious way at least to the case of a
gaussian variable with a physical boundary).

\item
\emph{I am not proposing the BMUA}!
(But those that think that the upper limit for $n=0$
should not depend on $b$
may consider the possibility of using the BMUA with
$\mu_{\mathrm{best}}^{\mathrm{min}} = 2$
instead of resorting to more complicated methods
that may even jeopardize the property of coverage\footnote{
By the way, I think that coverage is the most important property
of the Frequentist theory.
If coverage is not satisfied the results are statistically useless
in the contest of Frequentist theory.
}.)

\end{enumerate}

As shown in Fig.~\ref{aubelt},
the right edge of the confidence belt in the BMUA
is not very different from the one in the Unified Approach.
This is due to the fact that adding small values of $n$
with low probability to the acceptance intervals
has little effect.
Moreover,
it is clear that the acceptance interval for $\mu=0$
is equal for all Frequentist methods
with correct coverage that unify the treatment
of upper confidence limits
and two-sided confidence intervals.

\section{Bayesian Ordering}
\label{Ordering}

An elegant, natural and general way to obtain automatically
$\mu_{\mathrm{best}}^{\mathrm{min}} > 0$
is given by the
\emph{Bayesian Ordering} method
\cite{Giunti-bo-99},
in which
$\mu_{\mathrm{best}}$
is replaced by the Bayesian expectation value for $\mu$,
$\mu_{\mathrm{B}}$.

Choosing a natural flat prior,
the Bayesian expectation value for $\mu$
in a Poisson process with background
is given by
\begin{equation}
\mu_{\mathrm{B}}(n,b)
=
n + 1
-
\left(
\sum_{k=0}^n \frac{k b^k}{k!}
\right)
\left(
\sum_{k=0}^n \frac{b^k}{k!}
\right)^{-1}
=
n + 1
-
b
\left(
\sum_{k=0}^{n-1} \frac{b^k}{k!}
\right)
\left(
\sum_{k=0}^n \frac{b^k}{k!}
\right)^{-1}
\,.
\label{BEV}
\end{equation}
The obvious inequality
$
\sum_{k=0}^{n} k\,b^k/k!
\leq
n \sum_{k=0}^{n} b^k/k!
$
implies that
$\mu_{\mathrm{B}} \geq 1$.
Therefore,
the reference value for $\mu$ in the likelihood ratio
\begin{equation}
R(n,\mu,b)
=
\frac{ P(n|\mu,b) }{ P(n|\mu_{\mathrm{B}},b) }
=
\left( \frac{\mu+b}{\mu_{\mathrm{B}}+b} \right)^n
e^{\mu_{\mathrm{B}}-\mu}
\,,
\label{LR-BEV}
\end{equation}
that determines the construction of the acceptance intervals
as in the Unified Approach,
is bigger or equal than one.
As a consequence,
the decrease of the upper confidence limit
$\mu_{\mathrm{up}}$
for a given $n$
when the expected background $b$ increases
is significantly weaker than in the Unified Approach,
as illustrated in Fig.~\ref{higback}B.

Figure \ref{higback}C
shows $\mu_{\mathrm{up}}$
as a function of $b$ in the Bayesian Theory
with a flat prior and shortest credibility intervals\footnote{
In this case the posterior p.d.f. for $\mu$
is
\begin{equation}
P(\mu|n,b)
=
( b + \mu )^n \, e^{-\mu}
\left( \displaystyle n! \, \sum_{k=0}^{n} \frac{b^k}{k!} \right)^{-1}
\,,
\label{posterior}
\end{equation}
and the probability
(degree of believe)
that the true value of $\mu$ lies in the range
$[\mu_1,\mu_2]$
is given by
\begin{equation}
P(\mu\in[\mu_1,\mu_2]|n,b)
=
\left(
e^{-\mu_1} \sum_{k=0}^{n} \frac{(b+\mu_1)^k}{k!}
-
e^{-\mu_2} \sum_{k=0}^{n} \frac{(b+\mu_2)^k}{k!}
\right)
\left( \displaystyle \sum_{k=0}^{n} \frac{b^k}{k!} \right)^{-1}
\,.
\label{integral-probability}
\end{equation}
The shortest $100\alpha\%$ credibility intervals
$[\mu_{\mathrm{low}},\mu_{\mathrm{up}}]$
are obtained by choosing
$\mu_{\mathrm{low}}$
and
$\mu_{\mathrm{up}}$
such that
$P(\mu\in[\mu_{\mathrm{low}},\mu_{\mathrm{up}}]|n,b)=\alpha$
and
$P(\mu_{\mathrm{low}}|n,b) = P(\mu_{\mathrm{up}}|n,b)$
if possible
(with $\mu_{\mathrm{low}} \geq 0$),
or
$\mu_{\mathrm{low}} = 0$.
}.
One can see that the behavior of
$\mu_{\mathrm{up}}$
obtained with the Bayesian Ordering method is intermediate
between those
in the Unified Approach
and in the Bayesian Theory.
Although
one must always remember that the statistical meaning of
$\mu_{\mathrm{up}}$
is different in the two Frequentist methods
(Unified Approach and Bayesian Ordering)
and in the Bayesian Theory,
for scientists using these upper limits
it is often irrelevant how they have been obtained.
Hence, I think that an approximate agreement between
Frequentist and Bayesian results is desirable.

\begin{figure}[t]
\begin{minipage}[t]{0.48\textwidth}
\mbox{\includegraphics[bb=90 525 307 737,width=\textwidth]{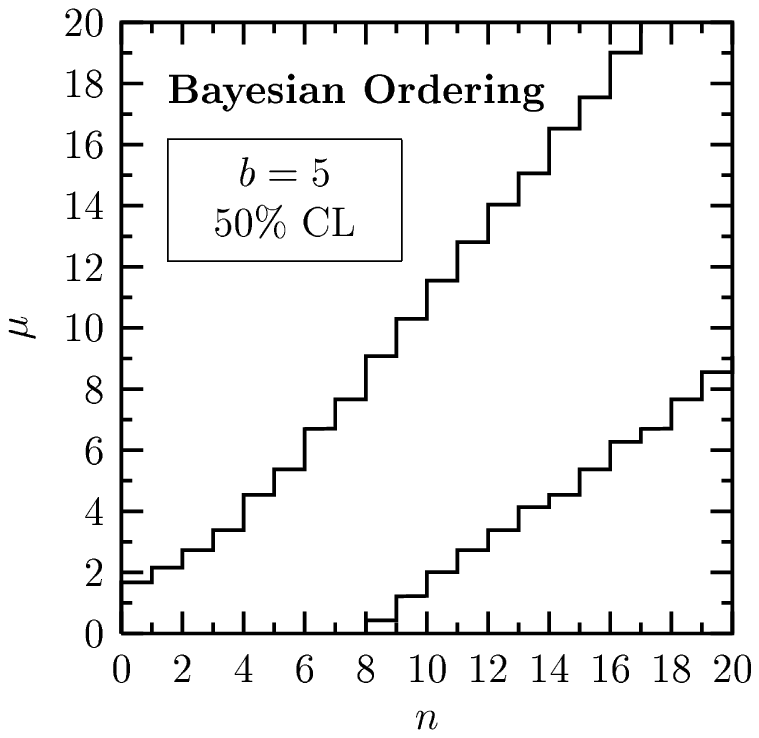}}
\caption{ \label{bolim}
Confidence belt obtained with the Bayesian Ordering
for background $b=5$ and confidence level $\alpha=0.90$.}
\end{minipage}
\hfill
\begin{minipage}[t]{0.48\textwidth}
\mbox{\includegraphics[bb=85 525 303 737,width=\textwidth]{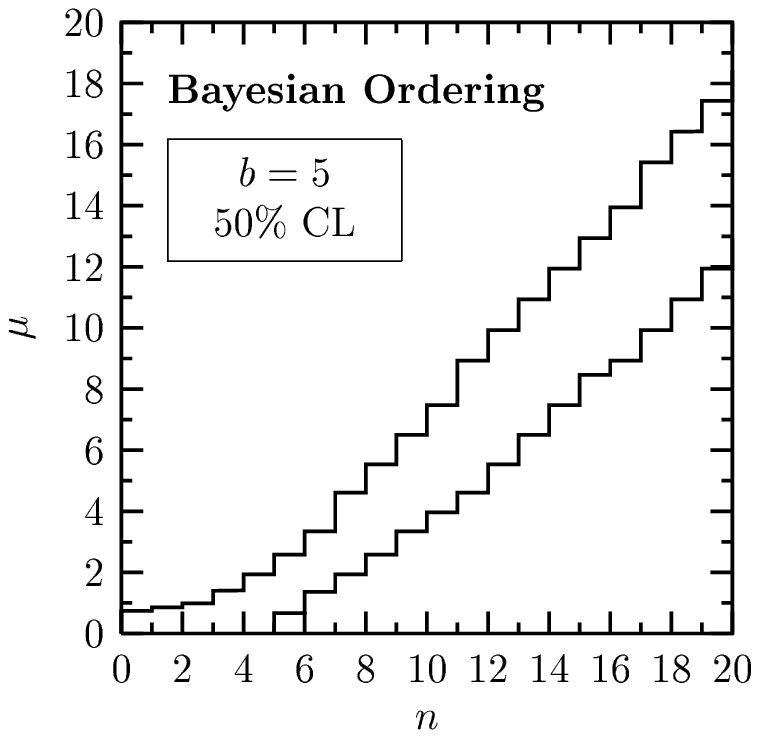}}
\caption{ \label{bobelt}
Confidence belt obtained with the Bayesian Ordering
for background $b=5$ and confidence level $\alpha=0.50$.}
\end{minipage}
\end{figure}

From Eq.~(\ref{BEV})
one can see that
\begin{eqnarray}
n \gg b
& \quad \Longrightarrow \quad &
\mu_{\mathrm{B}}(n,b)
\simeq
n + 1 - b
\simeq
n
\,,
\label{BEV-1}
\\
n \lesssim b
\,, \quad
b \gg 1
& \quad \Longrightarrow \quad &
\mu_{\mathrm{B}}(n,b)
\simeq
1
\,.
\label{BEV-2}
\end{eqnarray}
Therefore,
for $n \gg b$
the confidence belt obtained with the Bayesian Ordering method
is similar to that obtained with the Unified Approach.
The difference between the two methods show up
only for $n \lesssim b$.
This is illustrated in Figs.~\ref{bolim} and \ref{bobelt},
that must be confronted with the corresponding Figures
\ref{ualim} and \ref{uabelt}
in the Unified Approach.
Notice that,
as shown in Fig.~\ref{bobelt},
contrary to the Unified Approach,
the Bayesian Ordering method
gives physically significant (non-zero-width) confidence intervals
even for low values of the confidence level $\alpha$.

\section{Answers to some criticisms}
\label{Answers}

\noindent\textbf{Criticism:}
\textit{Bayesian Ordering is a mixture of Frequentism and Bayesianism.
The uncompromising Frequentist cannot accept it.}

No! It is a Frequentist method.

Bayesian theory is only used for the
\emph{choice of ordering}
in the construction of the acceptance intervals,
that in any case is subjective
and
beyond Frequentism
(as, for example, the central interval prescription
or the Unified Approach method).
The Bayesian method for such a subjective choice
is quite natural.

If you belong to the Frequentist Orthodoxy
(sort of religion!)
and the word ``Bayesian''
gives you the creeps,
you can change the name ``Bayesian Ordering''
into whatever you like
and use its prescription for the construction of the acceptance intervals
as a successful recipe.

\noindent\textbf{Criticism:}
\textit{In the Unified Approach (and maybe Bayesian Ordering?)
the upper limit on
$\mu$ goes to zero for every $n$ as $b$ goes to infinity, so that a low
fluctuation of the background entitles to claim a very stringent limit
on the signal.}

This is not true!

\begin{figure}[t]
\mbox{\includegraphics[bb=54 491 517 662,width=\textwidth]{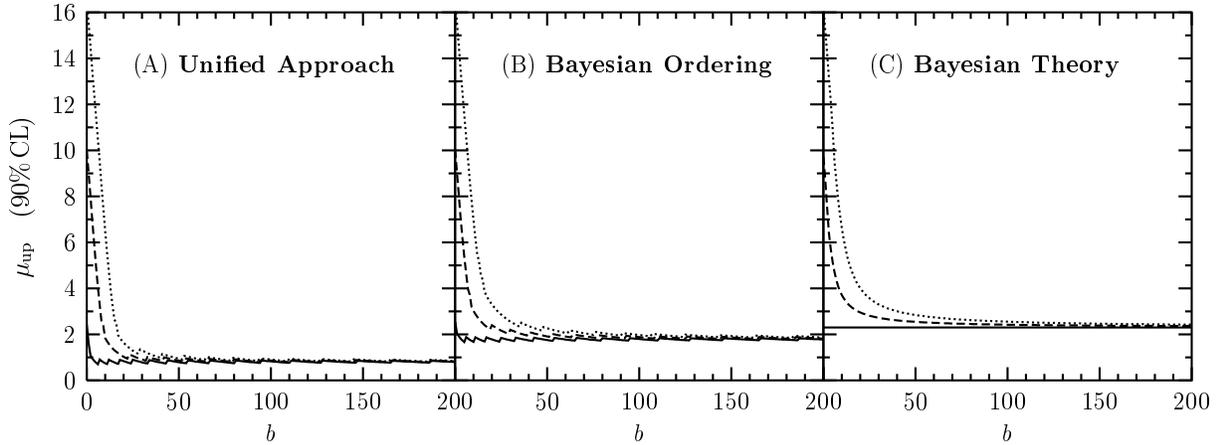}}
\caption{ \label{higback-asy}
90\% CL upper limit
$\mu_{\mathrm{up}}$
as a function of the background $b$
for
$n = 0$ (solid lines),
$n = 5$ (dashed lines)
and
$n = 10$ (dotted lines).
}
\end{figure}

One can see it\footnote{
In the Unified Approach
the likelihood ratio for $n \leq b$
is given by the expression in Eq.~(\ref{LR-1}),
that tends to $e^{-\mu}$
for $b \gg n$ and small $\mu$.
For $\mu\ll1$,
$e^{-\mu}\simeq1$
and all $n \ll b$
have
rank close to maximum.
For
$n > b$
the likelihood ratio is given by the expression in Eq.~(\ref{LR-2}).
For large values of $b$,
taking into account that $n > b$,
we have
$1 + \ln(\mu+b) - \ln{n} \simeq \ln b - \ln n < 0$
and
$\mu+b \simeq b$,
which imply that
$
R(n>b,\mu,b)
<
e^{-b}
\stackrel{b\to\infty}{\longrightarrow}0
$.
So the rank drops rapidly for $n>b$.
Therefore,
for small values of $\mu$
the $n$'s much smaller than $b$
have highest rank.
Since they have also very small probability,
they all lie
comfortably in the confidence belt,
if the confidence level $\alpha$
is sufficiently large
($\alpha \gtrsim 0.60$).
}
doing a calculation of the upper limit for
$\mu$ as a function of $b$ for large values of $b$.
The result of such a calculation
in the Unified Approach is shown in Fig.~\ref{higback-asy}A,
where the 90\% CL upper limit $\mu_{\mathrm{up}}$
is plotted as a function of $b$
in the interval $ 0 \leq b \leq 200 $
for $n=0$ (solid line), $n=5$ (dashed line) and $n=10$ (dotted line).
One can see that initially $\mu_{\mathrm{up}}$
decreases with increasing $b$,
but it stabilizes to about 0.8 for $b \gg n$,
with fluctuations due to the discreteness of $n$.
Figure~\ref{higback-asy}B
shows the same plot obtained with the Bayesian Ordering.
One can see that initially $\mu_{\mathrm{up}}$
decreases with increasing $b$,
but less steeply than in the Unified Approach,
and it stabilizes to about 1.8.
For comparison,
in Fig.~\ref{higback-asy}C
I plotted $\mu_{\mathrm{up}}$
as a function of $b$ in the Bayesian Theory
with a flat prior and shortest credibility intervals.
One can see that the behavior of $\mu_{\mathrm{up}}$
in the three methods considered in Fig.~\ref{higback-asy}
is rather similar.

\noindent\textbf{Criticism:}
\textit{For $n=0$ the upper limit $\mu_{\mathrm{up}}$
should be independent of the background $b$.}

But for $n>0$ the upper limit
$\mu_{\mathrm{up}}$
always decreases with increasing $b$!
It is true that for $n=0$
one is sure that no background event as well as no signal has been observed.
But this is just the effect of a low fluctuation of the background
that \emph{is present}!
Should we built a special theory for $n=0$?
I think that this is not interesting in the Frequentist framework,
because I guess that it leads necessarily to a violation of coverage
(that could be tolerated, but not welcomed,
only if it is overcoverage).

I think that
if one is so interested in having an upper limit $\mu_{\mathrm{up}}$
independent of the background $b$ for $n=0$,
one better embrace the Bayesian theory
(see Fig.~\ref{higback}C, Fig.~\ref{higback-asy}C
and
Ref.~\cite{Astone}),
which, by the way,
may present many other attractive qualities
(see, for example, \cite{DAgostini-YR3-99}).

\noindent\textbf{Criticism:}
\textit{A (worse) experiment with larger background $b$
should not give a smaller upper limit $\mu_{\mathrm{up}}$
for the same number $n$ of observed events.}

But, as shown in Fig.~\ref{higback},
this always happens!
Notice that it happens both for $n > b$ (dotted part of lines)
and for $n \leq b$ (solid part of lines),
in Frequentist methods as well as in the Bayesian Theory
(for $n>0$).
As far as I know,
nobody questions the
decrease of $\mu_{\mathrm{up}}$
as $b$ is increased if $n > b$.
So why should we question the same behavior when $n \leq b$?
The reason for this behavior is simple:
the observation of a given number $n$ of observed events
has the same probability if the background is small and the signal is large
or
the background is large and the signal is small.

I think that it is physically desirable
that and experiment with a larger background
do not give a \emph{much smaller} upper limit
for the same number of observed events,
but a \emph{smaller} upper limit is allowed
by \emph{statistical fluctuation}.
Indeed,
\begin{quote}
upper limits (as confidence intervals, etc.)
are statistical quantities that \emph{must fluctuate!}
\end{quote}
I think that the current race of experiments
to find the most stringent upper limit
is bad\footnote{
It is surprising that even at the Panel Discussion
\cite{Panel}
of this Workshop
(full of experts)
the statement
``the experimenters like to quote the smallest bound they can get away with''
was not strongly criticized.
What is the purpose of experiments?
(A) Give the smallest bound.
(B) Give useful and reliable information.
If your answer is (A) and you are an experimentalist,
I suggest that you stop deceiving us
and move to some more rewarding cheating activity.
},
because
it induces people to think that limits are fixed and certain.
Instead,
everybody should understand
that
\begin{quote}
a better experiment can sometimes give a worse upper limit
because of statistical fluctuations
and there is nothing wrong about it!
\end{quote}

\section{Conclusions}
\label{Conclusions}

In this report I have shown that
the necessity to choose a specific Frequentist method,
among several available,
does not introduce any degree of subjectivity
from a statistical point of view (Section~\ref{significance})
\cite{Giunti-Laveder-00}.
In other words,
all Frequentist methods are statistically equivalent.

However,
the physical significance of confidence intervals obtained
with different methods is different
and scientists interested in obtaining reliable and useful information
on the characteristics of the real world
must worry about this problem.
Obtaining empty or very small confidence intervals
for a physical quantity
as a result of a statistical procedure
is useless.
Sometimes it is even dangerous to present such results,
that lead non-experts in statistics
(and sometimes experts too) to false believes.

In Section~\ref{beauty}
I have discussed some virtues
and shortcomings of the Unified Approach
\cite{Feldman-Cousins-98}.
These shortcomings are ameliorated in the Bayesian Ordering method
\cite{Giunti-bo-99},
discussed in Section~\ref{Ordering},
that is natural, relatively easy,
and leads to more reliable upper limits.

In conclusion,
I would like to emphasize the following considerations:

\begin{itemize}

\item
One must always remember that,
in order to have coverage,
the choice of a specific Frequentist method
must be done
independently of the knowledge of the data.

\item
Finding some examples
in which a method fails does not
imply that it should not be adopted in the
cases in which it performs well.

\item
Since
all Frequentist methods are statistically equivalent,
\begin{center}
there is no need of a general Frequentist method!
\end{center}
In each case one can choose the method that works better
(basing the judgment on easiness, meaningfulness of limits, etc.).
Complicated methods with a wider range of applicability
are theoretically interesting,
but not attractive in practice.

\item
Somebody thinks that the physics community should agree on
a standard statistical method
(see, for example, \cite{Panel})\footnote{
As a theorist,
I find the argument,
presented by an experimentalist,
that a standard is useful
because otherwise one is tempted to analyze the data
with the method that gives the desired result
quite puzzling.
But if I were an experimentalist
I would be quite offended by it.
Isn't it a denigration of the professional integrity
of experimental physicists?
}.
In that case,
it is clear that this method must be always applicable.
But this is not the case, for example,
of the Unified Approach,
as shown in \cite{Zech}.
Although the Bayesian Ordering method has not
been submitted to a similar thorough examination,
I doubt that it is generally applicable.

I do not see
why experiments that explore different physics
and use different experimental techniques
should all use the same statistical method
(except a possible ignorance of statistics
and blind believe to ``authorities'').

I would recommend that
\begin{quote}
instead of wasting time on useless
characteristics as generality,
\emph{the physics community should worry about
the usefulness and credibility of experimental results}.
\end{quote}

\end{itemize}

\section*{Acknowledgements}

I would like to thank Marco Laveder for fruitful collaboration and
many stimulating discussions.

\end{document}